\documentclass[preprint]{revtex4-1}
\usepackage{graphicx}
\usepackage{natbib}
\usepackage{url}
\usepackage{graphicx,amssymb,amsmath}

\makeatletter

\newcommand{\Rmnum}[1]{\expandafter\@slowromancap\romannumeral #1@}
\makeatother

\begin{document}
\title{Thermal Properties of Disordered Li$_x$MoS$_2$ -- An Ab-Initio Study}

\author{Teut\"e Bunjaku}
\email{tbunjaku@iis.ee.ethz.ch}
\author{Mathieu Luisier}
\affiliation{Integrated Systems Laboratory, Department of Electrical
Engineering and Information Technology, ETH Zurich, Gloriastrasse
35, 8092 Zurich, Switzerland}

\begin{abstract}
An atomistic study of the thermal properties of lithiated molybdenum disulfide (MoS$_2$) is presented and an explanation for the experimentally determined anisotropic behavior of the in- and through-plane thermal conductivity is proposed. Configurations with different levels of lithium concentration are simulated using Density Functional Theory (DFT) and their structural, electronic, and thermal transport properties are evaluated as a function of the degree of lithiation. Comparisons between regular and disordered lithium distributions as well as to the experimental data reveal that the measured ratio of the in-plane to through-plane thermal conductivity can only be reproduced if the disorder of the lithium ions is taken into account.
These results suggest that it is not only possible to modify the thermal properties of MoS$_2$ by changing the degree of lithiation, but also by controlling its level of disorder.

\end{abstract}


\maketitle


\section{Introduction}
\begin{figure}
\includegraphics[width=0.75\textwidth]{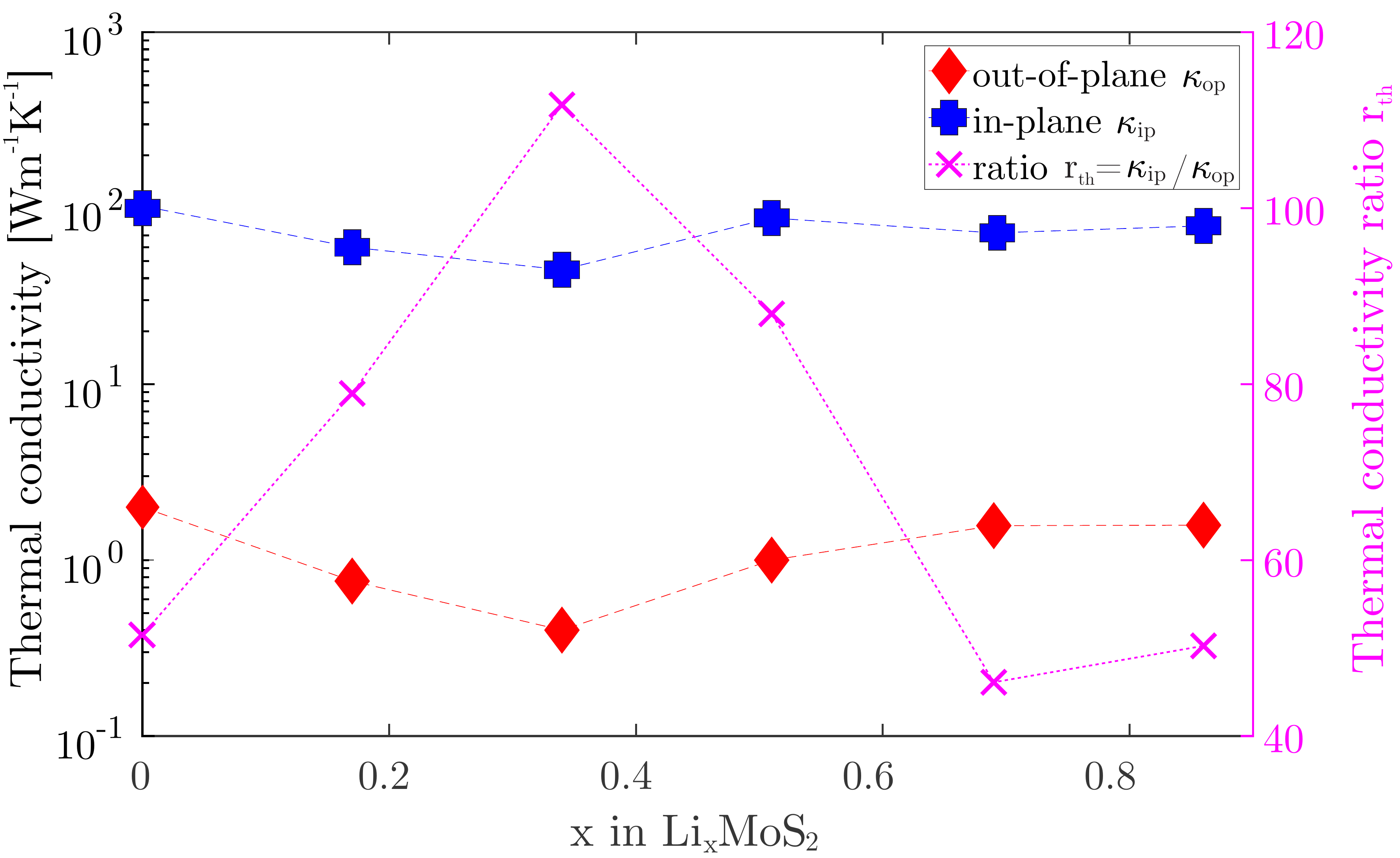}
\caption{Experimentally measured in-plane (crosses) and through-plane (diamonds) thermal conductivity of Li$_x$MoS$_2$ as a function of the lithiation level \cite{zhu2016tuning}. The right y-axis indicates the ratio (dashed line with x's) between both thermal conductivities, which increases upon lithiation until reaching its maximum at Li$_{0.34}$MoS$_2$ and then decreasing.}
\label{fig:paper}
\end{figure}

Motivated by the great promise of graphene in terms of electric, thermal, and mechanical properties \cite{balandin2011thermal}\cite{neto2009electronic}, other two-dimensional (2D) materials have recently started to attract a lot of attention from the scientific community \cite{choi2017recent}. Molybdenum disulfide (MoS$_2$), a semiconductor material part of the metal dichalcogenide family is structurally similar to graphite, which is composed of a planar layered structure. However, MoS$_2$ consists of triple atomic layers with strong intra-layer covalent bonds and weak layer-to-layer van der Waals forces. The large interlayer distance between the MoS$_2$ planes may be an ideal location to accommodate lithium (Li) ions for possible application as energy storage units \cite{hwang2011mos2}. It has also been observed that the intercalation of Li ions results in significant changes in the electronic and thermal properties of MoS$_2$, thus opening further avenues in the area of nanoelectronics \cite{wang2012integrated}\cite{cheng2014few}, optoelectronics \cite{choi2014lateral}, or thermoelectricity \cite{wan2015flexible}.\\
The thermal conductivity of lithiated bulk MoS$_2$ , Li$_x$MoS$_2$ ($0\leq x \leq 1$), was recently measured and it was shown that (i) it is anisotropic, (ii) the anisotropy between the in-plane $\kappa_{ip}$ and out-of-plane (through-plane) $\kappa_{op}$ components strongly depends on the Li concentration, and (iii) the ratio $r_{th}=\frac{\kappa_{ip}}{\kappa_{op}}$ reaches a maximum for Li$_{0.34}$MoS$_2$ \cite{zhu2016tuning}. These findings could be extremely important in electronic devices based on MoS$_2$ and other 2-D materials as an increased thermal conductivity is expected to enhance heat evacuation from their active region. In Figure \ref{fig:paper} the experimental out-of-plane as well as in-plane thermal conductivities of bulk MoS$_2$ are reported, together with the corresponding conductivity ratio r$_{th}$. Even though additional phonon modes are created by introducing Li ions into the MoS$_2$ structure, which should be beneficial to heat transport, the thermal conductivity first decreases along both directions as Li ions are intercalated between MoS$_2$ layers, as shown in Figure \ref{fig:paper}. This decrease is especially  surprising along the through-plane direction since the added Li ions bridge adjacent layers and are therefore expected to favor the propagation of phonons. \\
To take advantage of the relatively large modulations of the thermal properties of MoS$_2$, it is first necessary to understand the physics behind them. From the experiments, it could not yet been determined how the interplay between the intercalated Li ions and the MoS$_2$ layers affects the thermal conductivity of the formed material and why the intercalation process has a more pronounced influence on the out-of-plane direction. We therefore present here an ab-initio study of the thermal properties of Li$_x$MoS$_2$ to shed light on the underlying physical mechanisms and on the origin of the observed anisotropy. A phonon quantum transport solver approach relying on density functional theory (DFT) is used to do that and to provide insight into the structural, electronic, and thermal changes of MoS$_2$ upon lithiation.\\
For that purpose, we have evaluated the thermal properties of Li$_x$MoS$_2$ ($0\leq x \leq 1$) samples through thermal conductance/conductivity calculations at five different Li concentrations for regular and disordered Li placements. By comparing the results of the homogeneous (regular) and heterogeneous (disordered) structures, it can be deduced that the disposition of the Li ions strongly impacts the thermal properties of the composite material. The experimental measurements can only be qualitatively reproduced if a certain degree of disorder reigns in Li$_x$MoS$_2$. Structures with perfectly ordered Li ions in between two MoS$_2$ layers lead to an erroneous behavior of the $\kappa_{ip}$/$\kappa_{op}$ ratio, which continuously decreases as the Li concentration increases. With a carefully introduced disorder, our simulations indicate that the thermal anisotropy ratio grows upon lithiation until reaching its maximum at Li$_{0.25}$MoS$_2$, close to the experimental value of Li$_{0.34}$MoS$_2$, before starting to decrease. 
This confirms that the thermal properties of Li$_x$MoS$_2$ can be tuned by changing the Li concentration and the degree of  disorder. It should also be emphasized that the method implemented to calculate the thermal conductivity of lithiated MoS$_2$ can be applied to any other material with a disordered configuration.\\
The paper is organized as follows: first the simulation methodology is presented in Section \Rmnum{2} with an in-depth structural analysis of Li$_x$MoS$_2$ in subsection \textbf{A} and an overview of the corresponding thermal transport calculations in subsection \textbf{B}. The results are discussed in Section \Rmnum{3}, starting with the structural and electrical changes upon lithiation followed by the thermal properties in regular and disordered Li$_x$MoS$_2$ samples. A detailed comparison between the thermal conductivity of the homogeneous and heterogeneous structures is performed and a qualitative agreement with the experimental data from \cite{zhu2016tuning} is demonstrated. The paper ends with a conclusion in Section \Rmnum{4}, where the main results and findings are summarized. More details about the modeling approach are given in Appendix A.


\section{Methodology}
The structural and electronic properties of Li$_x$MoS$_2$ ($0\leq x \leq 1$) are extracted from density functional theory (DFT) as implemented in the Vienna Ab-Initio Software Package (VASP) \cite{kresse1996efficiency}\cite{kresse1996efficient}, whereas the dynamical matrices required to perform phonon transport are calculated using the finite displacement method implemented in phonopy \cite{phonopy}. The constructed structures and their corresponding dynamical matrices $\Phi$ allow us to compute the thermal current flowing through the considered atomic systems with the help of a quantum transport solver \cite{luisier2011investigation}\cite{luisier2006atomistic}. 
The whole process starts with the determination of stable atomic configurations for different Li concentrations within MoS$_2$ bulk samples. Their electronic properties are first calculated with DFT to gain insight into the structural and electronic changes upon lithiation. The produced stable structures are then used to simulate the thermal properties of Li$_x$MoS$_2$ ($0\leq x \leq 1$) with homogeneous and heterogeneous Li placement. The results are finally compared to each other and to experiments \cite{zhu2016tuning}.

\subsection{Structural Analysis}
The bulk phase of MoS$_2$ consists of molybdenum (Mo) sheets covalently bound to sulfur (S) atoms, forming planes that are held together by van der Waals forces. There
exist several metastable configurations of this material, as shown in previous experimental and theoretical works \cite{lin2016defect}\cite{voiry2015phase}. However, in this paper, we will mainly focus on two of them. The most stable atomic structure of bulk MoS$_2$ is the 2H phase, where each Mo atom has a trigonal prismatic coordination with its neighboring S atoms \cite{calandra2013chemically}, as illustrated in Figure 2(a). The 1T' phase is another metastable phase, which is achieved by slightly distorting the 1T configuration. In the latter, each Mo atom is octahedrally coordinated with its nearby S atoms. While the 1T phase is dynamically unstable for MoS$_2$, the 1T' can become stable under specific circumstances. Although the 1T' phase exhibits a lower total energy than the 1T one, it still has a bigger energy than the 2H atomic arrangement, which is characterized by an energy decrease of 0.576 eV per Mo atom. The bulk 2H and 1T' phases do not only differ in their structural properties, but also in their electronic ones, the 2H phase being a semiconductor with an indirect bandgap of 1.2 eV, while 1T' is metallic. \\
Due to its higher stability, the lithiation of MoS$_2$ is initiated in the 2H phase. As the Li concentration slowly increases, it becomes apparent that the intercalated atoms preferably sit at an octahedral site, binding to their neighboring S atoms. Upon lithiation, the material undergoes a phase transformation and changes from its 2H to 1T' configuration, resulting in not only electronic, but also thermal changes. It has been previously reported \cite{py1983structural} that the adsorption of Li atoms in the 1T' phase stabilizes the material, thus favoring a 2H-1T' phase transformation at x = 0.4 Li atoms per Mo. Both phases have been simulated here: periodic boundary conditions are applied along the x, y, and z directions of the MoS$_2$ samples to mimic bulk structures during the lithiation process. Supercells need to be used to account for the presence of disorder, as discussed later.\\
The energy and forces in Li$_x$MoS$_2$ are computed with VASP. For the structural calculations in the 2H phase, a k-point sampling resolution of 21x15x21 on a Monkhorst-Pack grid is used, whereas a 21x21x21 mesh is utilized for the 1T' phase. The generalized gradient approximation (GGA) of Perdew, Burke, and Ernzerhof (PBE) \cite{perdew1996generalized} is chosen as the exchange-correlation functional. The van der Waals forces are included through the  DFT-D2 method of Grimme \cite{grimme2006semiempirical}. Finally, the cutoff energy for the plane-wave basis is set to $550$~eV. For the structural end electronic calculations, convergence is considered to be achieved when the energy difference between two consecutive iterations of the self-consistent field is reduced below $10^{-6}$~eV and the force acting on each ion is smaller than 10$^{-3}$~eV/\AA. For thermal simulations, these values have to be further reduced ($10^{-9}$~eV and 10$^{-8}$~eV/\AA,~respectively) to ensure accurate phonon bandstructures without negative branches.

\subsubsection{2H phase - MoS$_2$ and Li$_{0.25}$MoS$_2$}
\begin{figure}
\includegraphics[width=0.9\textwidth]{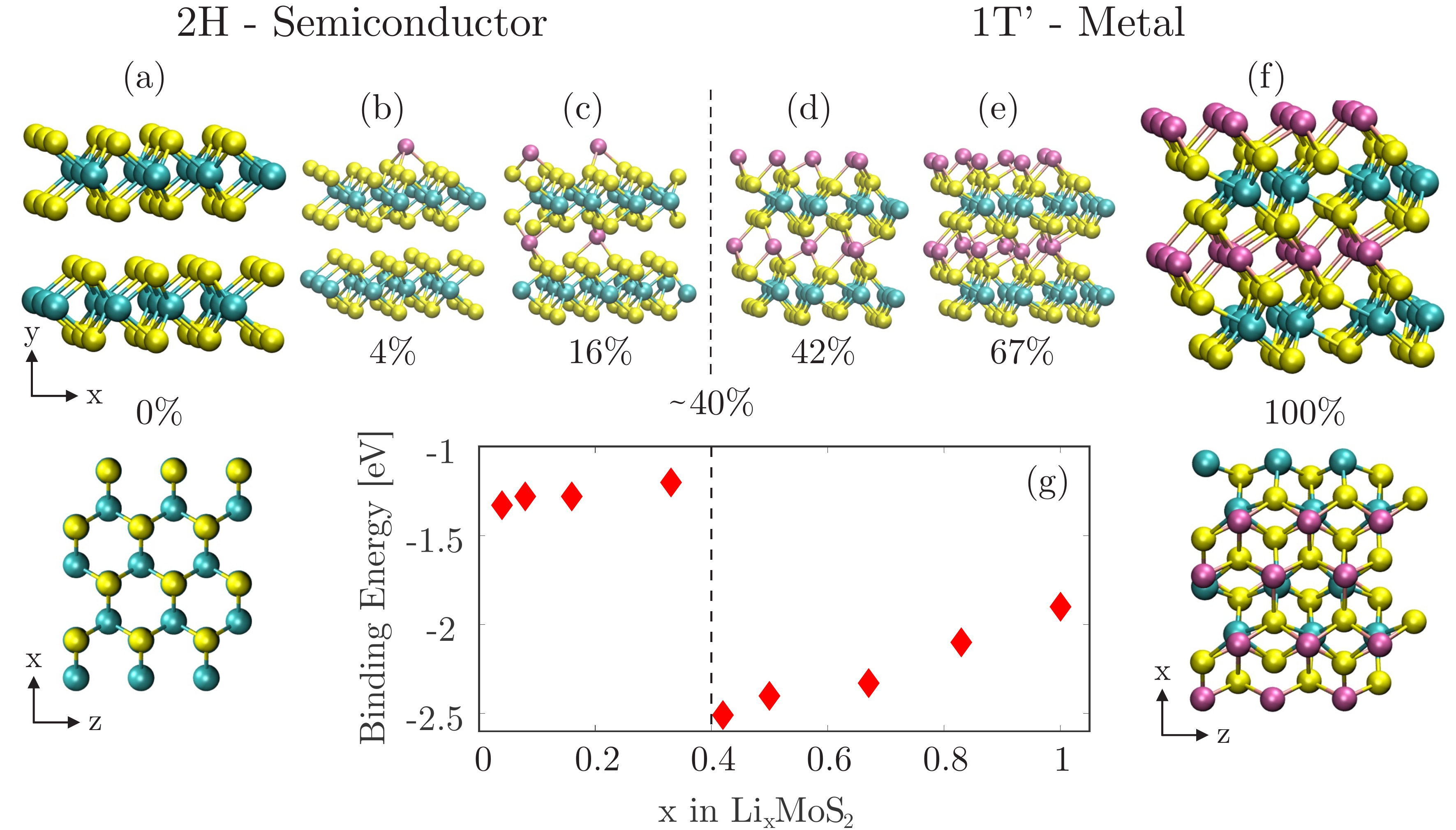}
\caption{(a-f) Structural changes of MoS$_2$ upon lithiation with the corresponding binding energy of the Li ions in MoS$_2$ (g). The yellow spheres represent Mo atoms, the red ones S atoms, and the lilac ones refer to Li atoms. The starting point is the semiconducting 2H phase of bulk MoS$_2$. By slowly adding Li atoms to the host material, the structure remains stable up to a 40\% concentration of Li, i.e. 0.4 Li atom per Mo. At this point the sample undergoes a phase transformation from 2H into the metallic 1T' phase and the Li binding energy abruptly drops. It should be noted that even at x=1, the host material has not reached its full Li capacity since the binding energy is still highly negative.}
\label{fig:binding}
\end{figure}

2H MoS$_2$ is built by considering the trigonal prismatic symmetry of the phase and its AB stacking behavior. Without Li ions in the host material, the considered unit cell is composed of 24 atoms (8 Mo, 16 S). The system can be cast into a rectangular supercell with the dimensions $x=5.52$~\AA, $y=12.41$~\AA, and $z=6.38$~\AA, where $x$ and $z$ are the in-plane directions and $y$ the out-of-plane one. In order to lithiate the 2H phase and reach a concentration of Li$_{0.25}$MoS$_2$, the total number of atoms in the unit cell must increase to 26 (2 Li, 8 Mo, 16 S). The added Li atoms most likely adhere to an octahedral site, where they are surrounded by their immediate S neighbors situated in the two most adjacent van der Waals layers, as can be seen in Figures \ref{fig:binding}(b) and \ref{fig:binding}(c). The dimensions of the rectangular supercell slightly change to $x=5.57$~\AA, $y=12.92$~\AA, and $z=6.43$~\AA$~$, suggesting a small expansion of the crystal caused by the lithiation process. The evolution of the lattice parameters along the three cartesian coordinates is plotted in Figure 3.\\

\subsubsection{1T' phase - Li$_{0.5}$MoS$_2$, Li$_{0.75}$MoS$_2$, and LiMoS$_2$}
After reaching a lithium concentration of 0.4~Li ions per MoS$_2$, the 2H phase evolves into the 1T' phase, which simultaneously affects the structural, electronic, and thermal properties of the crystal. Representative atomic structures are provided in Figure 4. The phase transition at x = 0.4 is very close to the value found experimentally (x=0.37) \cite{xia2017phase} and by other theoretical studies (x=0.4) \cite{calandra2013chemically}\cite{enyashin2012density}. Due to the altered symmetry and stacking (AA) of the van der Waals layers in the 1T' case, the total number of Mo and S atoms in the prepared rectangular unit cell can be reduced to 12. The need for two multilayers to properly represent the structure is not necessary anymore. For all concentrations in the 1T' phase, the number of Mo and S atoms is equal to 4 and 8, respectively, whereas the number of Li atoms increases from 2 at Li$_{0.5}$MoS$_2$, to 3 at Li$_{0.75}$MoS$_2$, and 4 at LiMoS$_2$. It is important to realize that the rectangular cell size is influenced by the degree of lithiation. This feature will be further discussed in the next Sections. 

\subsection{Thermal Properties}
Experimentally, the thermal conductivity $\kappa_{th}$ of Li$_x$MoS$_2$ was measured as a function of the Li concentration \cite{zhu2016tuning}. In the modeling effort, the thermal conductance $G_{th}$ is considered for device structures with a regular Li arrangement. A disorder-limited thermal conductivity $\kappa_{th}$ can be extracted for simulation domains with a random placement of Li ions between the MoS$_2$ layers.
Before calculating the thermal current flowing through a structure and evaluating its thermal conductance and conductivity, it is necessary to relax the unit cell with a high accuracy. A dynamical matrix can then be extracted using the finite displacement method. By knowing the dynamical matrix of the relaxed cell, it is possible to calculate the thermal current propagating through the material with the help of a quantum transport solver based on the Non-Equilibrium Green's Function (NEGF) formalism. To do that, a temperature gradient $\Delta T = 1$K is applied between both ends of the simulated samples. Such calculations can be performed for different device lengths and Li-ion concentrations. 
The thermal properties are first calculated for a regular Li placement at five different concentrations, namely Li$_x$MoS$_2$ with $x=(0, 0.25, 0.5, 0.75, 1)$. The results of regularly lithiated device structures are later compared to the results with disordered Li arrangements. The thermal current through disordered structures has been determined for five Li concentrations as well, i.e. $x=(0.125, 0.25, 0.5, 0.75, 0.875)$, replacing the perfectly ordered $x=(0,1)$ structures by components with $x=(0.125, 0.875)$. Note that a thermal conductivity could also be calculated for the $x=(0,1)$ cases, but this would require the inclusion of anharmonic phonon interactions, which is out of the scope of this paper. The whole simulation approach is now described in details step by step.\\

\subsubsection{Structure Relaxation and Dynamical Matrix Calculation}
After establishing stable configurations of Li$_x$MoS$_2$ based on a structural analysis, the produced geometry is relaxed with a more strict convergence criterion to eliminate spurious phonon bands characterized by negative frequencies. The energy difference between two consecutive self-consistent iterations should not exceed $10^{-9}$~eV and the force acting on each ion must be smaller than 10$^{-8}$~eV/\AA. These operations have to be done for each unit cell corresponding to a different Li concentration and Li distribution within the cell. The dynamical matrix is then calculated by creating a rectangular supercell and applying the finite displacement method \cite{phonopy} to obtain the second derivative of the harmonic potential. The size of the supercell and the k-point sampling strongly depend on the symmetry and the degrees of freedom of the selected unit cell. Typical supercell sizes range from $(x, y, z) = (13.25\mathrm{\AA}, 12.44\mathrm{\AA}, 11.83\mathrm{\AA})$ to $(19.89\mathrm{\AA}, 24.87\mathrm{\AA}, 17.74\mathrm{\AA})$ and their corresponding sampling schemes from $(k_x, k_y, k_z) = (3,3,3)$ to $(1,1,1)$.\\
In the case of regular Li placement, only one representative unit cell and its associated dynamical matrix have to be prepared for each concentration since it is assumed that the overall structure is homogeneous and made of repeated identical cells. For disordered Li$_x$MoS$_2$ crystals with Li concentrations $x=(0.25, 0.5, 0.75)$, all possible unit cell configurations exhibiting the desired Li concentration have to be built and randomly grouped with each other to give rise to meaningful simulation domains. To obtain intermediate concentrations, e.g. $x=0.125$ (or 0.875), all possible MoS$_2$ and Li$_{0.25}$MoS$_2$ (or Li$_{0.75}$MoS$_2$ and LiMoS$_2$) configurations have to be randomly mixed with each other until reaching the desired Li level. Atomic arrangements for different Li concentrations are shown in Figure \ref{fig:all_struc}. Each of them refers to a rectangular unit cell that can be either repeated (homogeneous case) or combined with others (disordered case) to form a device structure with a given Li concentration. The resulting systems are simulated with the aid of a quantum transport solver \cite{luisier2006atomistic}. The detailed procedure of constructing the dynamical matrices with a random placement of Li ions is outlined in Appendix A.

\subsubsection{Quantum Transport Solver - Thermal current}
Using VASP and phonopy a dynamical matrix with periodic boundary conditions in all three directions is obtained for the rectangular unit cells. Before being able to use it in a quantum transport solver, it has to be put into a finite, device-like form. This can be done by first slightly modifying the dynamical matrix and removing interactions exceeding a pre-defined cutoff radius $r_{cut}$. This parameter is critical since crucial information may be lost if the value is set to small. At the same time, the computational burden exponentially increases with it. To ensure a proper representation of the bandstructure after the dynamical matrix has been modified, it is important to compare it to the reference bandstructure extracted from phonopy \cite{phonopy}. Such a comparison is provided in Figure \ref{fig:cut} for 2H MoS$_2$ with the original phonon bands shown as solid red lines and the bands resulting from the truncated dynamical matrix represented by dashed blue lines. As can be seen, both phonon dispersions agree well with each other if a cut-off radius $r_{cut}=0.7nm$ is chosen. For all considered structures, the maximum difference between the original and modified bands is not allowed to exceed 10\%, thus ensuring accurate quantum transport calculations. \\
The next step consists in assembling a larger device structure made of the small rectangular unit cells that have been prepared. This process is repeated for all investigated Li concentrations. For the regular configurations, the simulated super-structures can be created by repeating the same rectangular unit cell several times along the transport direction until reaching the desired length. The dynamical matrix is built accordingly, i.e. by constructing a block tri-diagonal matrix where all diagonal ($\Phi_{00}$), upper ($\Phi_{01}$), and lower ($\Phi_{10}$) blocks are identical and derived from the same rectangular unit cell, as schematically illustrated in the upper half of Figure 6. Note that the block $\Phi_{00}$ contains all inter-atomic interactions within a rectangular unit cell, while $\Phi_{01}$ ($\Phi_{10}$) represents the coupling with its image cell situated at $x_0+\Delta$ ($x_0-\Delta$). The position $x_0$ refers to the location of the unit cell along the x-axis and $\Delta$ to the width of the unit cell along the same axis.\\
The simulation domains for disordered Li placements are created using two different methods. For Li$_x$MoS$_2$ with $x=(0.25, 0.5, 0.75)$ the device structures are built by randomly stringing together the different cell configurations with the same Li concentration, but different atomic arrangement. For example, for Li$_{0.25}$MoS$_2$ , there exist four rectangular unit cells with the targeted Li level. They can be randomly mixed with each other to form a proper device structure. A schematic of this process is shown in the lower half of Figure 6. The dynamical matrix of the produced sample has to follow the same order as the super-structure built from different rectangular unit cells. For $x=(0.125, 0.875)$ the approach has to be slightly adapted since the structures and their dynamical matrices have not been directly computed at these concentrations. The thermal properties of Li$_{0.125}$MoS$_2$ (Li$_{0.875}$MoS$_2$) can still be calculated by randomly stacking all possible Li$_{0.25}$MoS$_2$ (Li$_{0.75}$MoS$_2$) and MoS$_2$ (LiMoS$_2$) configurations together, which gives the correct Li concentration. Again, the dynamical matrix has to follow the same order as the device structure, each different rectangular unit cell being associated to a specific matrix block, as detailed in Appendix A. \\
After setting up the simulation domains and the corresponding dynamical matrices of both regular and disordered situations, the thermal current flowing through them can be determined as well as the resulting thermal conductance and conductivity as a function of the Li concentration. This can be done by solving the following quantum transmitting boundary method (QTBM) problem \cite{lent1990quantum}
\begin{equation}
	(\omega^2-\bf{\Phi}-\bf{\Pi}^{RB})\bf{\varphi} = Inj,
\end{equation}
where $\omega$ is the phonon frequency, $\bf{\Phi}$ the DFT-based dynamical matrix of the whole system built as in Eq. (A.1) to (A.3) in Appendix A, $\bf{\Pi}^{RB}$ the retarded boundary self-energy, $\bf{\varphi}$ a vector containing the crystal vibrations along all cartesian coordinates, and \textbf{Inj} an injection vector \cite{rhyner2013phonon}. The self-energy and the injection vector represent the open boundary conditions. They allow phonons to enter and exit the simulation domain \cite{luisier2011investigation}. Determining $\bf{\Pi}^{RB}$ and \textbf{Inj} is done as in reference [14] for electrons, the Hamiltonian and dynamical matrices having both a similar sparsity pattern. This formulation of the quantum transport problem is equivalent to the NEGF formalism, but computationally more efficient because it takes the form of a sparse linear system of equations "Ax=b". The assumed periodic y and z directions are discretized with N$_{qy}$=31 and N$_{qz}$=33 phonon momentum points. From $\varphi$, a transmission function $T_{ph}(\omega)$ can be calculated, which leads to the thermal current $I_{th}$ when the Landauer-B\"uttiker formalism is applied \cite{rhyner2013phonon}\cite{markussen2009electron}. For the heterogeneous structures, it is further possible to compute the disorder-limited thermal conductivity by first determining the length-dependent thermal current $I_{th}(L)$ at each Li concentration and then computing the corresponding thermal resistance $R_{th}(L)$
\begin{equation}
R_{th}(L) = \frac{\Delta T}{I_{th}(L)},
\end{equation}
where $\Delta T$ is the temperature difference between the device left and right contacts and L the length of the simulation domain along the transport direction (x). From the thermal resistance, the thermal conductivity can be derived by applying the dR/dL method \cite{rim2002low}, assuming diffusive transport
\begin{equation}
	\kappa_{th}=\frac{\Delta L}{\Delta R_{th}} = \frac{L_{max}-L_{min}}{R_{th}(L_{max})-R_{th}(L_{min})}.
\end{equation}
Here, $\Delta L$ is the difference between the longest ($L_{max}$) and shortest ($L_{min}$) device and $\Delta R$ the corresponding thermal resistance variation. To account for the influence of randomness, 50 different samples have been simulated per Li concentration and device length. The thermal conductivity has been averaged over them. Typical simulation results and the validation of the chosen approach are reported in Figure \ref{fig:conductance_resistance}. The linear increase of $R_{th}(L)$ vs. $L$ indicates diffusive transport, as needed to apply Eq. (3).


\section{Results and Discussion}

The simulated structural, electronic, and thermal properties of lithiated bulk MoS$_2$ have been compared to the experimental data of Zhu et al. \cite{zhu2016tuning}. We started with the 2H phase of bulk MoS$_2$ and slowly increased the Li concentration until reaching LiMoS$_2$ and its 1T' phase. This gave us a comprehensive overview of the underlying intercalation processes and their consequences on the behavior of Li$_x$MoS$_2$. The key importance of the disordered Li placement will be highlighted in the following. It is the main reason behind the evolution of the ratio between the in-and through-plane conductivity vs. the Li concentration. Neither the simple presence of Li atoms between MoS$_2$ planes nor the geometrical changes caused by intercalation can explain the occurrence of a ratio maximum at around $x=0.34$ \cite{zhu2016tuning}.

\subsection{Structural and Electrical Changes}
\begin{figure}
\includegraphics[width=0.75\textwidth]{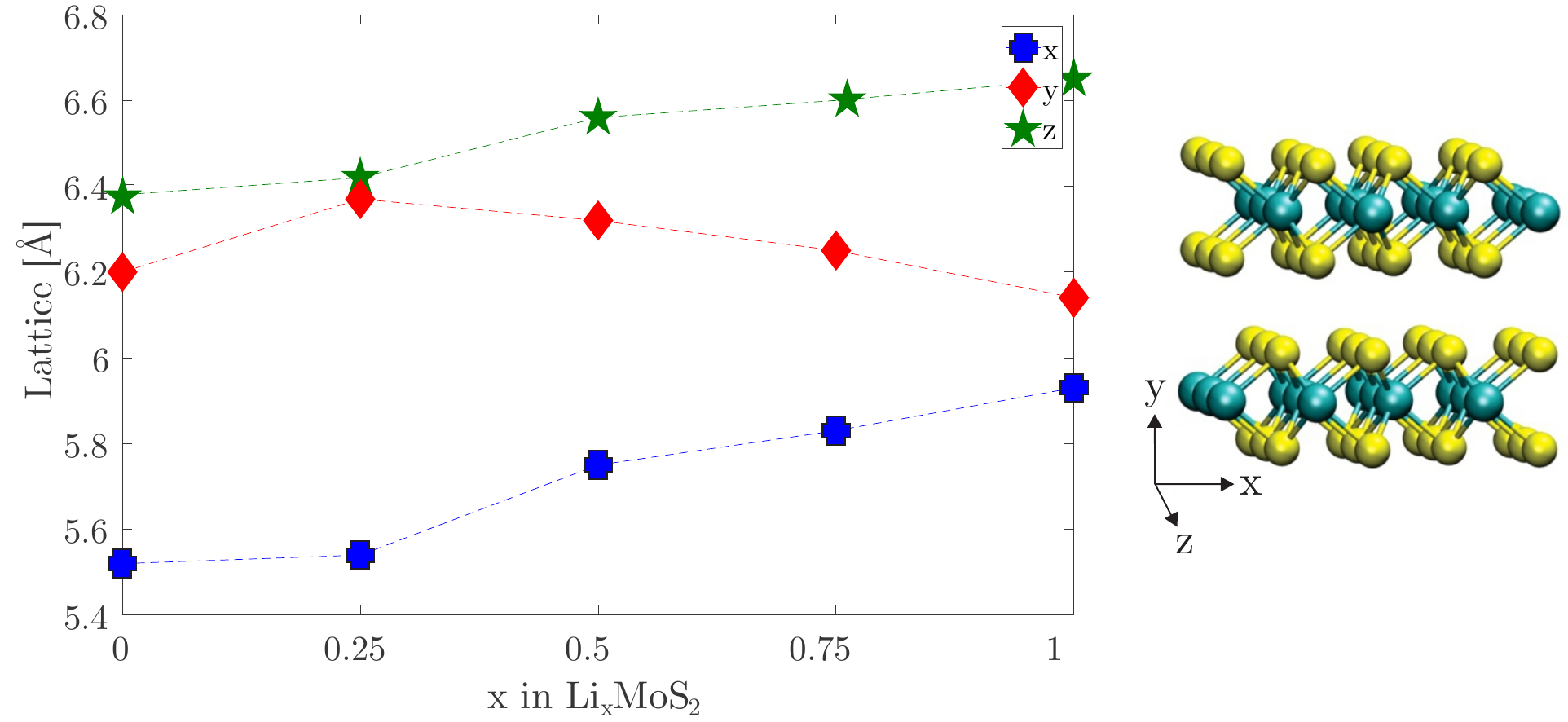}
\caption{Changes of the lattice parameters of the rectangular Li$_x$MoS$_2$ unit cell upon lithiation. The in-plane lattices x (crosses) and z (stars) continuously increase upon lithiation, while the through-plane lattice y (diamonds) first increases in the 2H phase and then decreases after the 2H-1T' phase transformation.}
\label{fig:lattice}
\end{figure}

\begin{figure}
\includegraphics[width=0.7\textwidth]{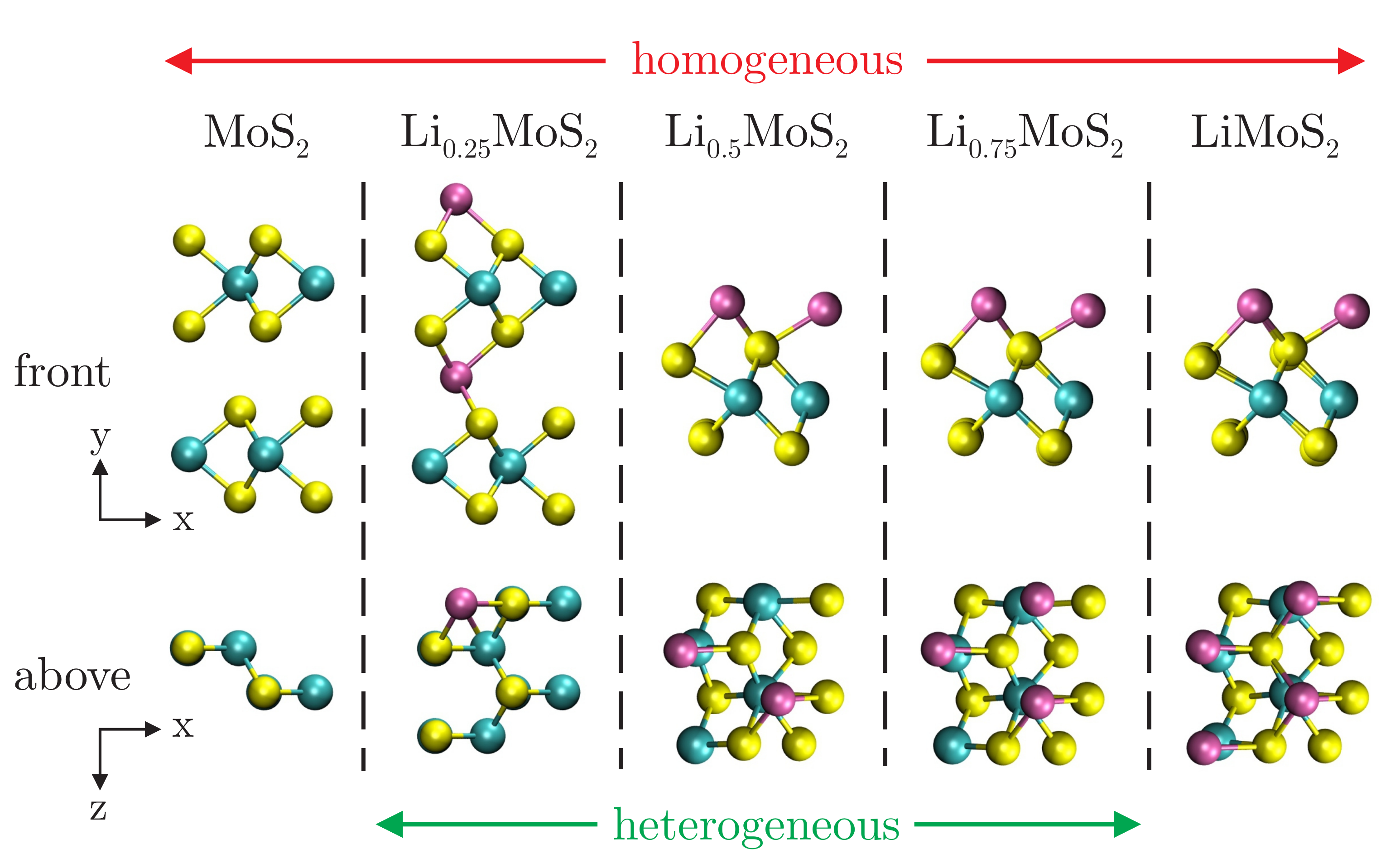}
\caption{Representative rectangular Li$_x$MoS$_2$ unit cells seen from the front (top row) and from above (bottom row). While for x=0 and 1, there exists only one single atomic configuration, 4 different unit cells can be created for x=0.25, 6 for x=0.5, and 4 for x=0.75, while keeping the total number of atoms per cell constant.}
\label{fig:all_struc}
\end{figure}

\begin{figure}
\includegraphics[width=0.7\textwidth]{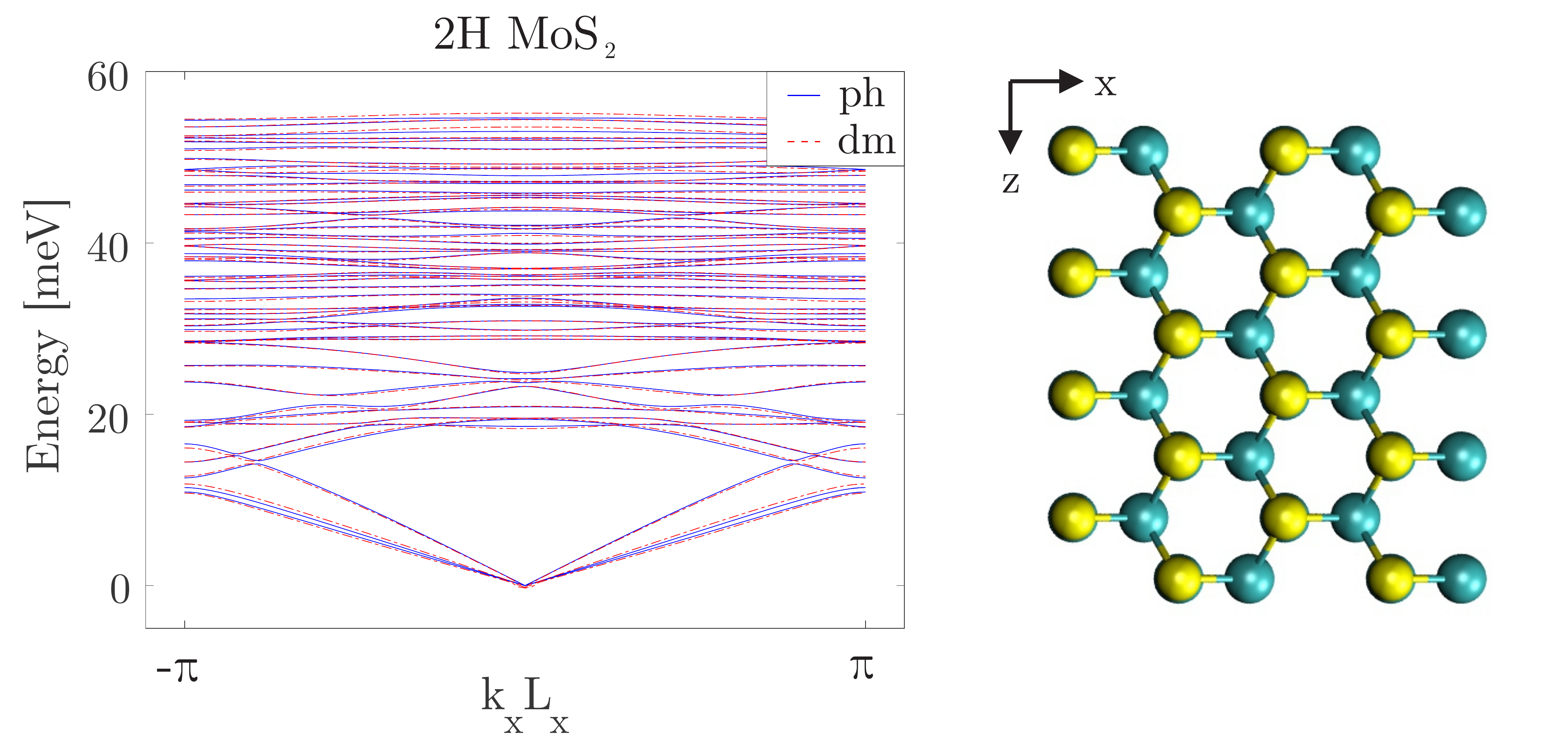}
\caption{Phonon dispersion (left) of the 2H MoS$_2$ unit cell shown on the right as directly calculated with phonopy (ph, solid lines) and after introducing a cut-off radius $r_{cut}=7.0$\AA~ (dm, dashed lines) into the dynamical matrix and importing it to a quantum transport solver.} 
\label{fig:cut}
\end{figure}

Adding Li atoms to the 2H phase of bulk MoS$_2$ at different positions makes it possible to determine the most favorable binding sites of the intercalated components in this material. While increasing the Li concentration, the Li binding energy should be monitored to ensure that the capacity of the host material is not exceeded. Until reaching 0.4 Li per MoS$_2$ the Li binds at the octahedral sites with an energy of $-1.35$~eV, as shown in Figure \ref{fig:binding}(g). After Li$_{0.4}$MoS$_2$ the crystal undergoes a phase transformation into the metallic 1T' configuration. In this case, the binding energy first decreases abruptly to a value of $-2.5$~eV for Li$_{0.42}$MoS$_2$ and then continually increases until reaching $-1.9$~eV for LiMoS$_2$. Even after reaching LiMoS$_2$, the Li capacity of the host material is not fully reached. The possibility to go beyond x=1 makes MoS$_2$ especially appealing for battery applications. Similar to the 2H phase, in the 1T' case, the added Li atoms adhere to an octahedral site, where they bind to their neighboring S atoms. \\
While the Li concentration increases and the phase of the material transforms, the lattice parameters change as well, as indicated in Figure \ref{fig:lattice}. The through-plane lattice first increases in the 2H phase from 6.2\AA~at MoS$_2$ to 6.37\AA~at Li$_{0.25}$MoS$_2$ and then decreases to 6.14\AA~at LiMoS$_2$. The in-plane lattice parameters both slightly increase during the whole process from 5.52\AA~in the x direction in MoS$_2$ to 5.93\AA~in LiMoS$_2$, and from 6.38\AA~to 6.65\AA~along the z direction. The overall changes in the lattice parameters, especially in the through-plane direction are too small to fully explain the behavior of the thermal conductivity and its anisotropic properties. Hence, other mechanisms have to be investigated to shed light on the experimental data.\\

\subsection{Thermal Properties - Regular Lithium Arrangement}
Using the structures depicted in Figure \ref{fig:all_struc}, the thermal conductance has been calculated at five different Li loading levels. For each of these five concentrations, the Li atoms are arranged in a perfectly regular manner throughout the simulated domains, making them ideal configurations. The thermal conductance results for the through- and in-plane directions are plotted in Figure \ref{fig:conductance}. It can be seen that the overall behavior does not coincide with the experimentally measured values from Figure \ref{fig:paper} and reference \cite{zhu2016tuning}. \\
In our simulations, the out-of-plane thermal conductance $G_{op}$ first slightly decreases until reaching Li$_{0.25}$MoS$_2$, before drastically increasing to a value two times larger at Li$_{0.5}$MoS$_2$ than it was at Li$_{0.25}$MoS$_2$. It then continues increasing until reaching its maximum at x=1. In the in-plane direction, the conductance $G_{ip}$ does not vary much: it slightly decreases until Li$_{0.5}$MoS$_2$ and then re-increases. Note that since the two in-plane thermal conductances along x and z are very similar, we will be showing only the values for the x direction. Putting everything together, the corresponding thermal conductance ratio $G_{ip}/G_{op}$continuously decreases with increasing Li concentration, contradicting the experiments, where a maximum of this ratio at around $x=0.34$ can be observed. This clearly indicates that the intercalation of the Li ions into the MoS$_2$ and the corresponding lattice changes alone cannot explain the experimental data. What is missing is the possibility for Li ions to be randomly placed between two MoS$_2$ layers. Since the degree of disorder depends on the Li concentration, the thermal conductance ratio is significantly affected by the presence of randomness.

\subsection{Thermal Properties - Disordered Lithium Arrangement}

Adding disorder to the five concentration levels gives and extra layer of complexity and makes the model more realistic. The thermal current of Li$_x$MoS$_2$ samples with \\$x=(0.125, 0.25, 0.5, 0.75, 0.875)$ can be computed at different lengths. From the resulting length-dependent thermal resistance, the thermal conductivity can be derived. For that purpose, 50 different configurations are considered for each of the five Li concentrations and four lengths between $30$ and $60$~nm. As expected, the thermal resistance linearly increases from $L=30$ to $60$~nm, as demonstrated in Figure \ref{fig:conductance_resistance} for $x=(0.25,0.5,0.75)$. The same happens for $x=(0.125, 0.875)$. When transport is diffusive, as above, the $dR/dL$ method in Equation (3) can be applied to obtain the thermal conductivity $\kappa$ along the in- and out-of-plane directions. The calculated thermal conductivities are shown in Figure 9. It can be seen that the qualitative behavior of the simulated data agrees well with the experimental measurements, contrary to the results with regular Li arrangements. Our ab-initio quantum transport calculations predict a maximum of the in-plane/through-plane thermal conductivity ratio for $x=0.25$, very close to the experimental value of $x=0.34$. From this comparison, it can be concluded that the added disorder dominates the behavior of the thermal conductivity and fully determines the ratio between $\kappa_{ip}$ and $\kappa_{op}$. It should be noted that the case $x=0.375$ could not be computed since this configuration requires the mixing of 2H (Li$_{0.25}$MoS$_2$) and 1T' (Li$_{0.5}$MoS$_2$) unit cells. The large structural differences make the mixing of these two phases unreliable.\\
The simulated results further highlight that the propagation of the thermal current is 18 times more favorable in the in-plane than the through-plane direction at Li$_{0.25}$MoS$_2$. The intercalation of Li ions thus provides an efficient way to tune the thermal conductivity of MoS$_2$ and the anisotropy of the in- and out-of-plane directions. Such feature could find applications in thermoelectricity. It could also be used to improve the evacuation of heat from the active region of Li ion batteries or to realize a thermal transistor.\\
It should be emphasized that anharmonic phonon scattering was not included in our calculations. In spite of that, the experimental data are well reproduced, except that our thermal conductivity values are lower than the experimental one. These discrepancies could possibly be attributed to structural differences between simulations and experiments: in our calculations, disorder is probably much more local than in experiments where regions with very different Li concentrations co-exist, but only the average over these regions is reported. Furthermore, it should not be excluded that in case of disorder, anharmonic phonon interactions can in fact (slightly) increase the thermal conductivity by connecting different phonon branches \cite{luisier2013thermal}.

\section{Conclusion}
The structural, electronic, and thermal properties of bulk Li$_x$MoS$_2$ (0$\leq$x$\leq$1) have been investigated using first-principles method. The structural and electronic properties could be determined with the help of ground state calculations, while the thermal conductance/conductivity was investigated with the aid of a quantum transport solver. Upon lithiation, a phase transformation of MoS$_2$ has been observed at a concentration of 0.4 Li atom per Mo species, which agrees well with previous experimental and theoretical data. After the phase transformation, the initial semiconducting 2H phase is replaced by a metallic 1T' configuration.\\
Calculating the thermal conductance of regularly lithiated structures at five different Li concentrations revealed that available experimental results, especially the ratio between the in-plane and through-plane thermal properties, could not be captured by such a modeling approach. This discrepancy could be overcome by adding disorder to the  Li$_x$MoS$_2$ samples, making them more realistic. An original method to construct the dynamical matrix of disordered structures and extract their thermal properties was therefore proposed and tested on lithiated MoS$_2$. Comparing the thermal conductivity of disordered configurations with the experimental data resulted in qualitatively similar results.  
The introduced disorder plays an essential role in Li$_x$MoS$_2$, causing the thermal conductivity ratio to increase while in the 2H phase and to decrease after its phase transformation into 1T'. These results suggest that the degree of disorder directly affects the thermal conductivity of 2-D materials. By changing the concentration and disorder of the intercalated atoms, it is possible to engineer the thermal properties of the host material.\\

\begin{acknowledgments}
This research has been funded by the EU Commission under the ERC Starting Grant: E-MOBILE. The computer simulations have been done at the Swiss National Supercomputer Center under projects s662 and pr28 (supported by PRACE).
\end{acknowledgments}



%

\appendix
\section{Dynamical matrix of systems with disordered Li distributions}
Since DFT tools can generally not be used to create the dynamical matrix of large systems, two methods have been established to do that for our Li$_x$MoS$_2$ systems with disorder. The idea consists in randomly assembling small rectangular unit cells to form a supercell, which then builds our simulation domain.\\ 
First, we consider the case Li$_x$MoS$_2$ with $x=(0.25, 0.5, 0.75)$. The diagonal blocks of the dynamical matrix, $\Phi_{ii}^{aa}$ are directly taken from the phonopy calculations \cite{phonopy} of the corresponding rectangular unit cell of type T$_a$. The procedure is more complex for the off-diagonal blocks $\Phi_{ij}^{ab}$ that connect a unit cell of type T$_a$ located at $x=x_i$ to another one of type T$_b$ at $x=x_j$. Since the positions of the Mo and S atoms remain the same in both cells, the matrix entries describing these interactions do not have to be altered. However, the values of the interactions between Li-Mo, Li-S, and Li-Li situated in two adjacent cells of type T$_a$ and T$_b$ must be adjusted to guarantee a proper coupling of the cells. To account for that, the $\Phi_{ij}^{ab}$ blocks are assembled as follows:

\begin{equation}
\Phi_{ij}^{ab} =
  \begin{cases}
    \frac{\Phi_{01}^{aa}+ \Phi_{01}^{bb}}{2} & : \text{Mo-Mo, S-S, Mo-S, S-Mo},\\
    \Phi_{01}^{bb}                          & : \text{Mo-Li, S-Li},\\
    \Phi_{01}^{aa}                          & : \text{Li-Mo, Li-S},\\
  \end{cases}
\end{equation}
where the $\Phi_{01}^{mm}$ blocks connect a rectangular unit cell of type T$_m$ to its next, identical neighbor along x. These blocks can also be directly derived from phonopy calculations \cite{phonopy}. For the Li-Li interactions between two cells of type T$_a$ and T$_b$, the situation is more complicated and distinctions between several cases have to be made:
\begin{equation}
\Phi_{ij}^{ab} =
  \begin{cases}
   0 & : d_{ab}\geq r_{cut}\\
    \Phi_{01}^{aa}                          & : 0.95\cdot d_{aa} \leq d_{ab} \leq 1.05\cdot d_{aa}~ \text{AND}~ d_{ab} \geq 1.1\cdot d_{bb}\\
    \Phi_{01}^{bb}                          & : 0.95\cdot d_{bb} \leq d_{ab} \leq 1.05\cdot d_{bb}~ \text{AND}~ d_{ab} \geq 1.1\cdot d_{aa}\\
    10\cdot \frac{\Phi_{01}^{aa}+\Phi_{01}^{bb}}{2}                          & : d_{ab} \approx 0.5\cdot d_{aa}~ \text{AND}~ d_{ab}\approx 0.5\cdot d_{bb}\\
       \frac{\Phi_{01}^{aa}+\Phi_{01}^{bb}}{2}                          & : \mathrm{otherwise}\\
  \end{cases}
\end{equation}
where $d_{aa}=||\text{Li}^a-\text{Li}^a||$ ($d_{bb}=||\text{Li}^b-\text{Li}^b||$) is the distance between two Li atoms situated in two identical neighboring cells T$_a$ (T$_b$), $d_{ab}$ is the distance between two Li atoms located one in cell T$_a$, the other in T$_b$, and $r_{cut}$ is the cut-off radius behind which the interactions are set to zero. The factor 10 for the fourth case could be empirically determined when comparing the entries of two dynamical matrices, where in one of them the distance between the Li atoms has been halved.   
The $\Phi_{ji}^{ab}$ blocks can be built by transposing the $\Phi_{ij}^{ab}$. To ensure that this approach does not affect the thermal current flowing through Li$_x$MoS$_2$ devices, a validation test has been performed. A quantum transport simulation domain has been created using two different cells a and b. Its dynamical matrix has then been constructed using the equations above. An identical structure has been built with a larger unit cell c composed of the blocks a and b. Both components have been repeated in transport direction, as shown in Figure \ref{fig:test_hetero}. The calculated thermal current through the repeated a+b and c devices agree very well, with a difference of less than 5\%, as does the energy-resolved transmission function. This indicates that the chosen method to compute the dynamical matrix of disordered systems is not only computationally more efficient (smaller blocks needed in DFT), but also accurate. Hence, it has been used all along the paper . \\
To construct device structures for Li$_{0.125}$MoS$_2$ and Li$_{0.875}$MoS$_2$, the approach has to be slightly adapted. The diagonal blocks $\Phi_{ii}^{aa}$ can be directly taken from the phonopy calculation \cite{phonopy} of the rectangular unit cell of type T$_a$, as before. The off-diagonal block $\Phi_{ij}^{ab}$ connecting a unit cell of type T$_a$ (for example Li$_{0.25}$MoS$_2$) with another one of type T$_b$ (for example MoS$_2$) is created by applying the following rules:
\begin{equation}
\Phi_{01}^{ab} =
  \begin{cases}
    \frac{\Phi_{01}^{aa}+ \Phi_{01}^{bb}}{2} & : \text{Mo-Mo, S-S, Mo-S, S-Mo, Mo-Li$_{ab}$, Li$_{ab}$-Mo, S-Li$_{ab}$, Li$_{ab}$-S, Li$_{ab}$-Li$_{ab}$},\\
    \Phi_{01}^{bb}                          & : \text{Mo-Li$_b$, S-Li$_b$},\\
    \Phi_{01}^{aa}                          & : \text{Li$_a$-Mo, Li$_a$-S},\\
  \end{cases}
\end{equation}
where Li$_{ab}$ represent Li atoms that are present in both cell types T$_a$ and T$_b$, whereas Li$_a$ (Li$_b$) are Li atoms that can only be found in the T$_a$ (T$_b$) rectangular cell.
For the interactions between the Li atoms the same method as in Equation (A2) can be used.

\newpage

\begin{figure}
\begin{center}
\includegraphics[width=0.85\textwidth]{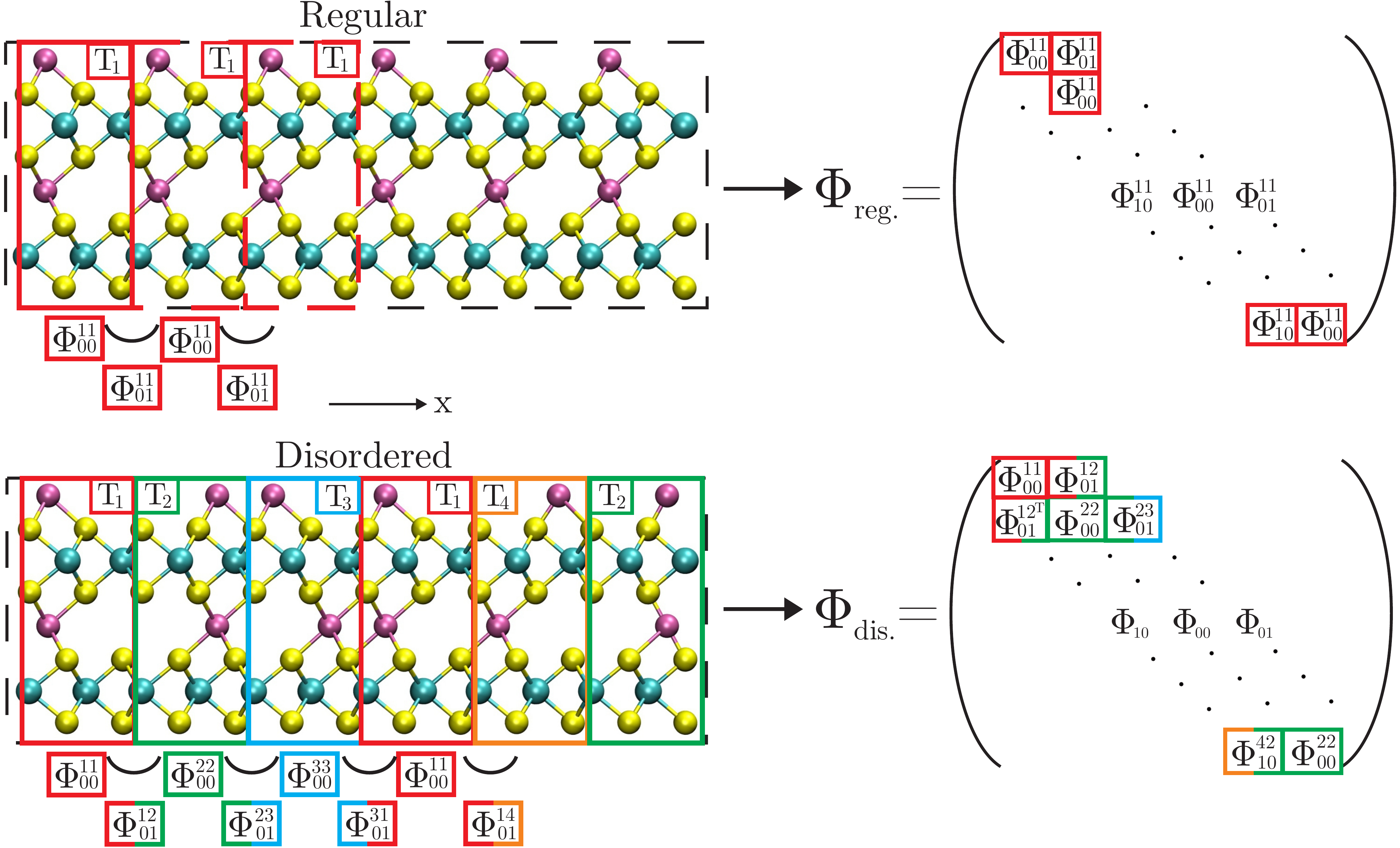}
\caption{Device structure (left) and corresponding dynamical matrix (right) for regular (upper row) and disordered (lower row) Li$_{0.25}$MoS$_2$ configurations. In the regular case, the same rectangular unit cell (T$_1$) is repeated along the transport direction x and the dynamical matrix $\Phi_{reg.}$ is built accordingly. In case of disorder, rectangular unit cell with the same Li concentration, but different atomic arrangements are grouped together. Four types of rectangular unit cells must be considered for $x=0.25$. They are labeled T$_1$ to T$_4$. The construction of the overall dynamical matrix $\Phi_{dis.}$ requires a special treatment of the cell coupling matrices $\Phi_{01}^{ij}$ and $\Phi_{10}^{ij}$, where $i$ and $j$ can be different and comprised between 1 and 4.}
\end{center}
\label{fig:hamiltonian}
\end{figure}

\begin{figure}
\includegraphics[width=0.6\textwidth]{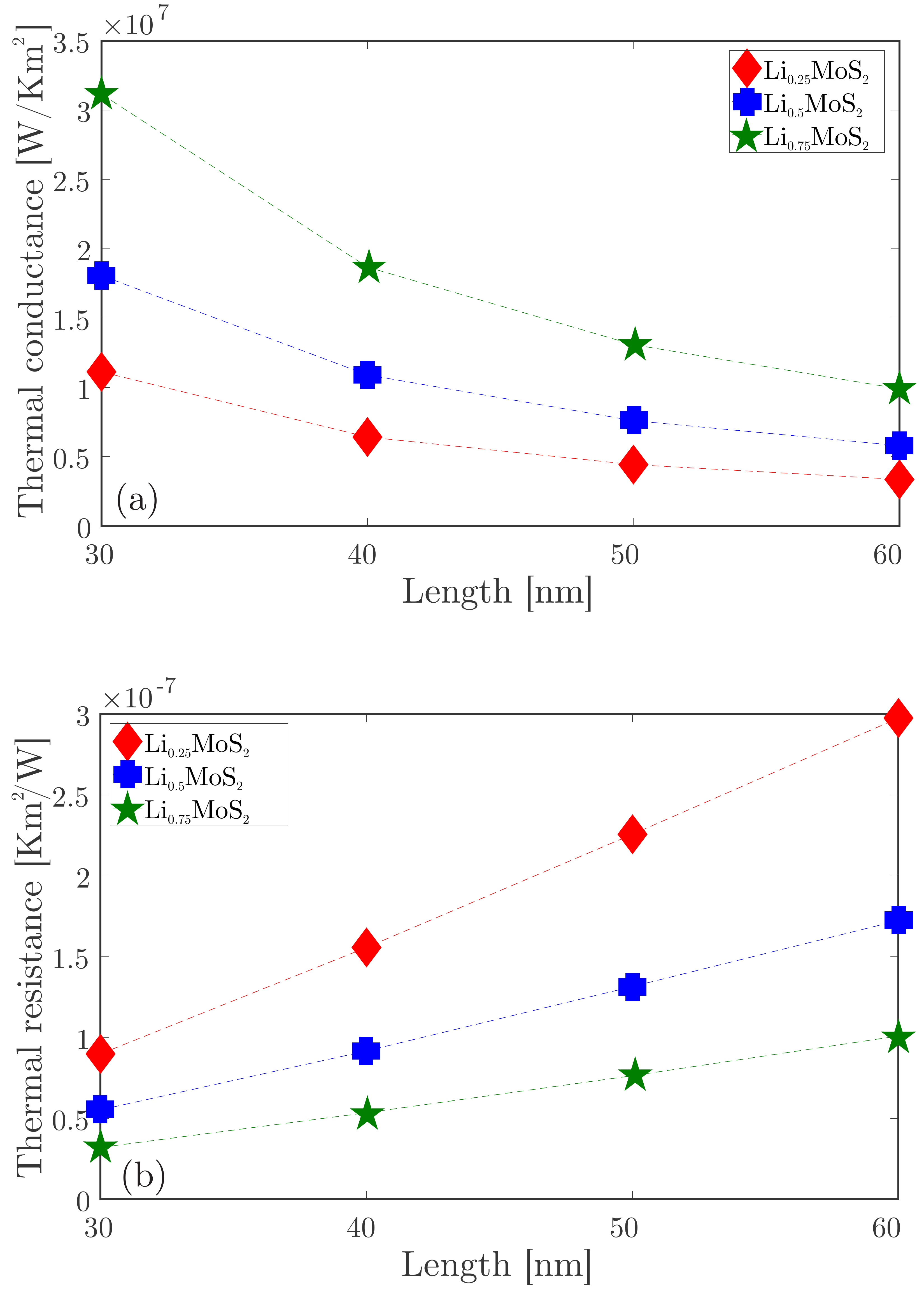}
\caption{(a) Thermal conductance G$_{th}$ of Li$_x$MoS$_2$ with $x=(0.25, 0.5, 0.75)$ as a function of the sample length along the transport direction x. (b) Corresponding thermal resistance $R_{th}=1/G_{th}$. Each point represents the average of 50 different configurations. As expected, $R_{h}$ linearly increases with the sample length (diffusive transport).}
\label{fig:conductance_resistance}
\end{figure}

\begin{figure}
\includegraphics[width=0.7\textwidth]{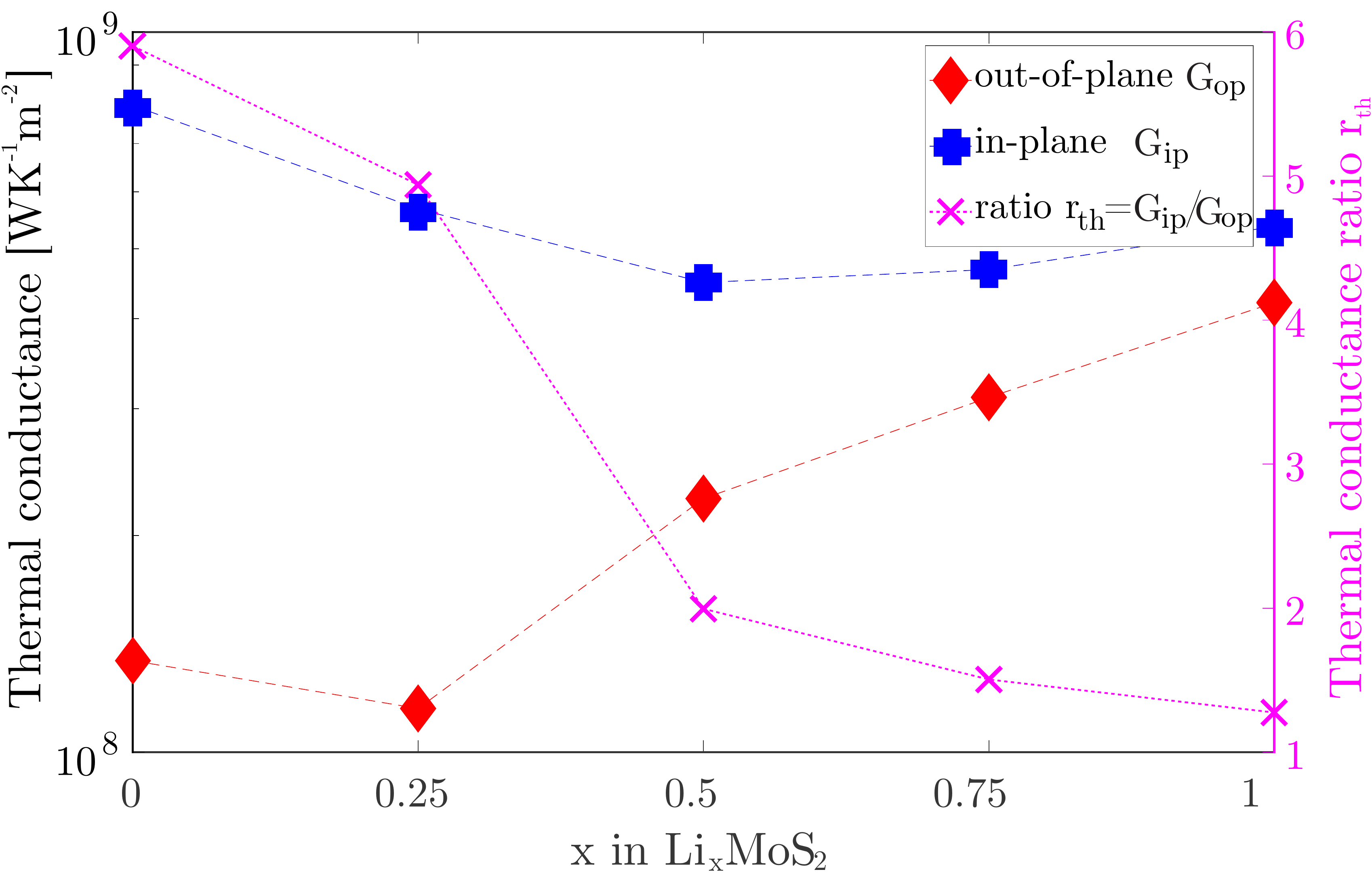}
\caption{Thermal conductance of Li$_x$MoS$_2$ as a function of the Li loading level x for perfectly ordered structures. The line with crosses refers to the in-plane conductance $G_{ip}$, the line with crosses to the out-of-plane value $G_{op}$, whereas the line with x's represents the ratio between them.}
\label{fig:conductance}
\end{figure}

\begin{figure}
\includegraphics[width=0.7\textwidth]{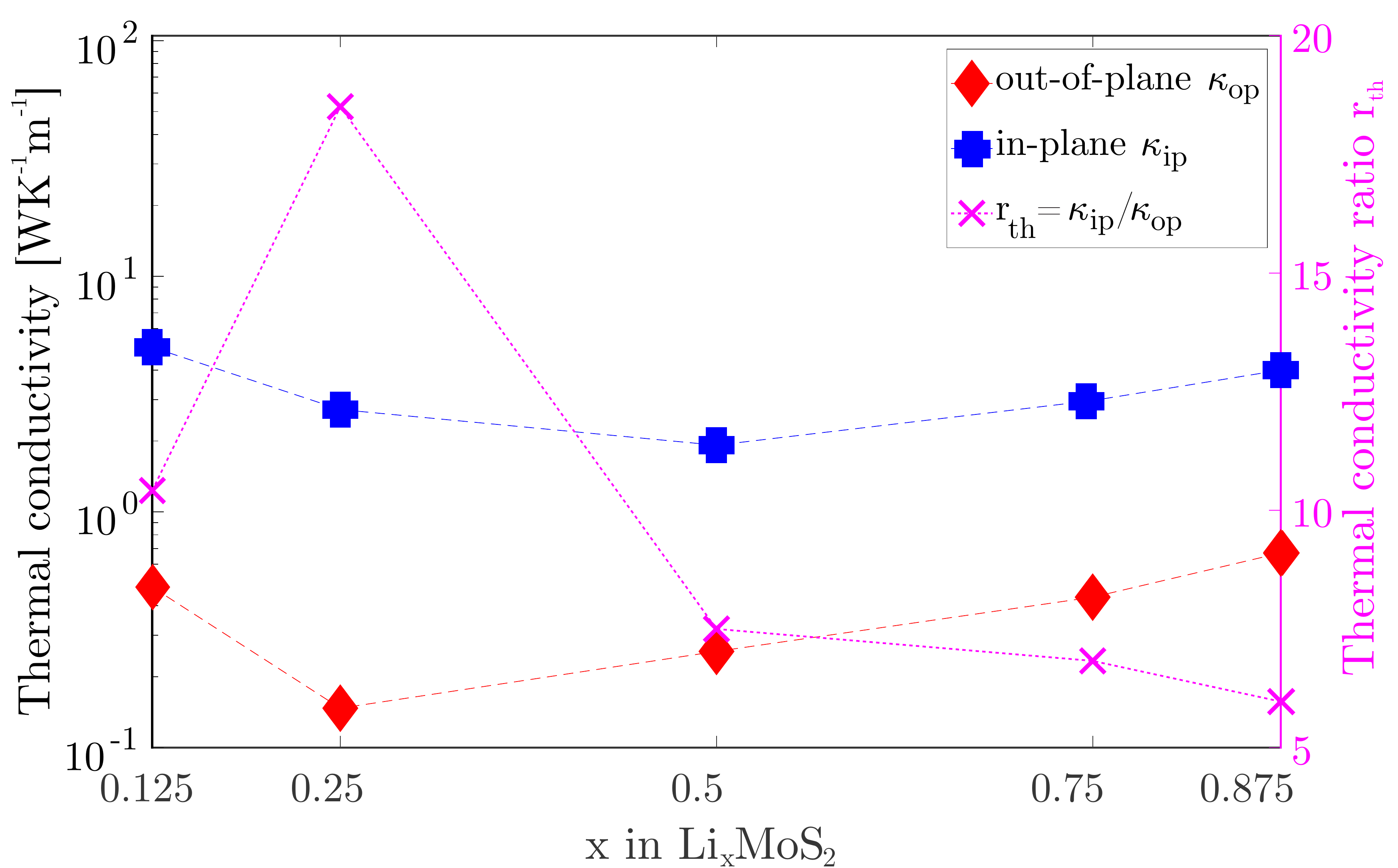}
\caption{Thermal conductivity of Li$_x$MoS$_2$ with random placement of the Li ions as a function of the Li concentration $x=(0.125, 0.25, 0.5, 0.75, 0.875)$. The in-plane (dashed line with crosses) and out-of-plane (dashed line with diamonds) are plotted together with the ratio between them (line with x's).}
\label{fig:conductance_dis}
\end{figure}

\begin{figure}
\includegraphics[width=0.6\textwidth]{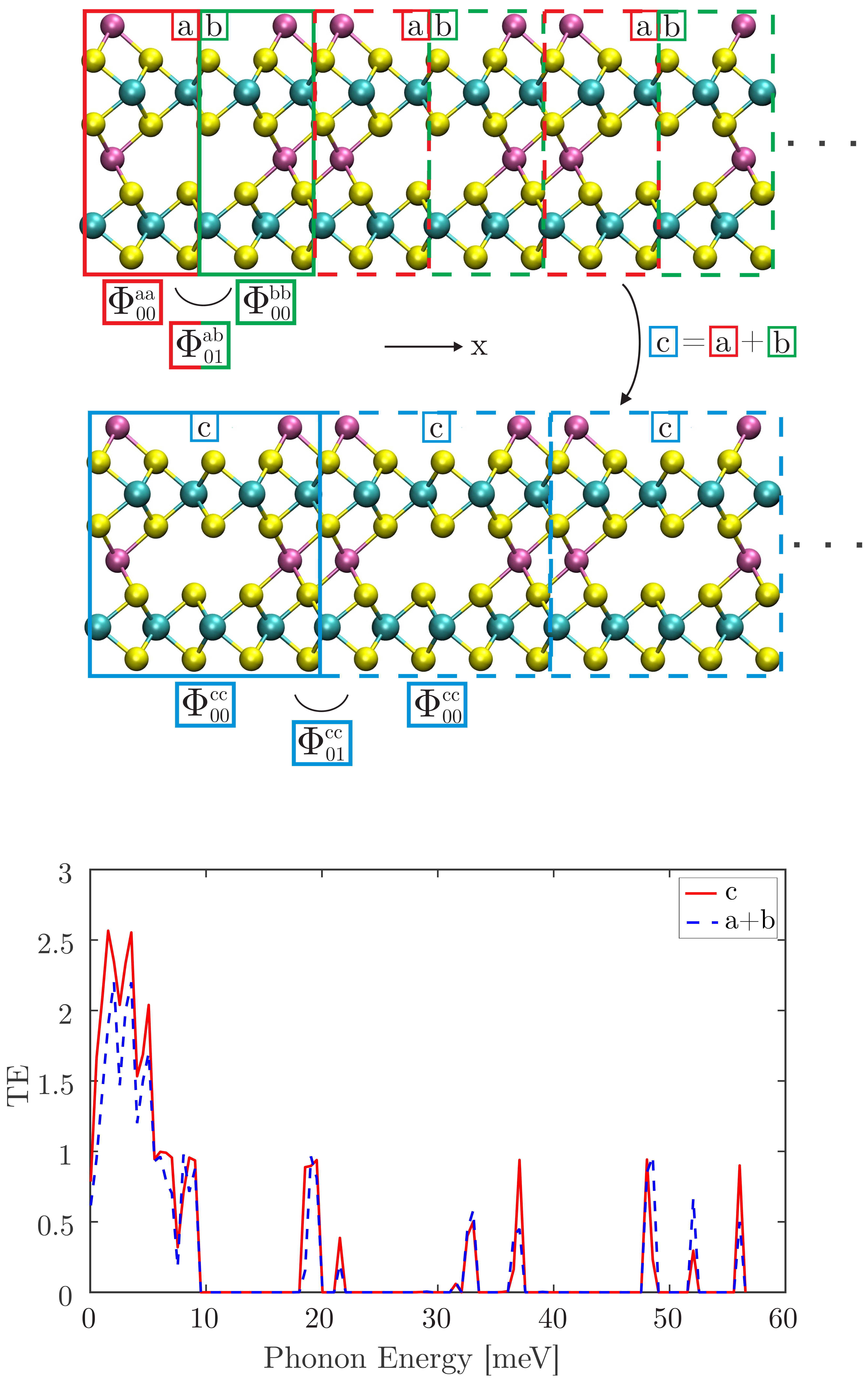}
\caption{(top) Test device structure used to validate the approach of Appendix A to construct the dynamical matrix of Li$_x$MoS$_2$ systems with random Li placement. In the upper plot, the dynamical matrices of two rectangular cells a (red) and b (green) are first created before they are coupled to each other with the help of Equations (A1)-(A2). In the lower part, the dynamical matrix of the supercell c=a+b is directly generated. Both the a+b and c (blue) cells are then repeated along the x-axis. (bottom) Energy-resolved transmission function through a sample made of the repetition of the a+b (blue dotted line) and c (red solid line) unit cells.}
\label{fig:test_hetero}
\end{figure}

\end{document}